\documentclass[letterpaper,structabstract]{aa}
\usepackage{graphicx}
\usepackage{txfonts}

\begin{document}

\def\deg{$^{\rm o}$}

\titlerunning{A search for diffuse radio emission in relaxed, cool-core galaxy clusters}

\title{A search for diffuse radio emission in the relaxed, cool-core galaxy clusters A1068, A1413, A1650, A1835, A2029, and Ophiuchus}
\author{F. Govoni\inst{1}
          \and
          M. Murgia\inst{1,2}
          \and
          M. Markevitch\inst{3}
          \and
          L. Feretti\inst{2}
         \and
          G. Giovannini\inst{2,4}
         \and
          G.B. Taylor\inst{5}
         \and 
          E. Carretti\inst{2}   
          }
\institute{
              INAF - Osservatorio Astronomico di Cagliari,
              Poggio dei Pini, Strada 54, I--09012 Capoterra (CA), Italy 
           \and   
              INAF - Istituto di Radioastronomia, 
              Via Gobetti 101, I--40129 Bologna, Italy
           \and
              Harvard-Smithsonian Center for Astrophysics, 60 Garden Street,
              Cambridge, MA 02138
           \and 
              Dipartimento di Astronomia, 
              Univ. Bologna, Via Ranzani 1, I--40127 Bologna, Italy
          \and
              University of New Mexico, MSC 07 4220, Albuquerque, NM 87131. Greg Taylor is also an Adjunct Astronomer at the National Radio Astronomy Observatory.
                }
\date{Received; accepted}

\abstract
{}
{ We analyze sensitive, high-dynamic-range, observations
to search for extended, diffuse, radio emission in relaxed and 
cool-core galaxy clusters.
}
{We performed deep 1.4 GHz Very Large Array observations, 
 of A1068, A1413, A1650, A1835, A2029, and complemented our dataset 
with archival observations of Ophiuchus.
}
{We find that, in the central regions of A1835, A2029, and Ophiuchus, 
the dominant radio galaxy is surrounded by
diffuse low-brightness radio emission that takes the form of a mini-halo.
We detect no diffuse emission in A1650, at a surface brightness level 
of the other mini-halos. 
We find low significance indications of diffuse emission in 
A1068 and A1413, although to be classified as mini-halos they
would require further investigation, possibly with data of
higher signal-to-noise ratio. 
In the Appendix, we report on the serendipitous 
detection of a giant radio galaxy with a total spatial extension 
of $\sim$1.6 Mpc.
}
{}

\keywords{Galaxies:clusters:individual:A1068, A1413, A1650, A1835, 
A2029, Ophiuchus - radio continuum: galaxies}

\offprints{F. Govoni, email fgovoni@ca.astro.it}
\maketitle


\section{Introduction}

There is now firm evidence that the intra-cluster medium (ICM)
consists of a mixture of hot plasma, magnetic fields, and relativistic 
particles. 
While the baryonic content of the galaxy clusters is dominated
by the hot ($T\simeq 2 - 10$ keV) intergalactic gas, whose thermal emission is
observed in X-rays, a fraction of clusters also exhibit megaparsec-scale 
radio halos
(see e.g., Feretti \& Giovannini 2008, Ferrari et al. 2008, and references
therein for reviews).
Radio halos are diffuse, low-surface-brightness 
($\simeq 10^{-6}$ Jy/arcsec$^2$ at 1.4 GHz), 
steep-spectrum\footnote{S($\nu$)$\propto \nu^{- \alpha}$} ($\alpha > 1$) 
sources,
permeating the central regions of clusters, produced by
synchrotron radiation of relativistic electrons with energies of
$\simeq 10$ GeV in magnetic fields with $B\simeq 0.5-1\;\mu$G.
Radio halos represent the strongest evidence 
of large-scale magnetic fields and relativistic particles
throughout the intra-cluster medium.

Radio halos are typically found in clusters that display
significant evidence for an ongoing merger 
(e.g., Buote 2001, Govoni et al. 2004). 
Recent cluster mergers were proposed to play an 
important role in the reacceleration 
of the radio-emitting relativistic particles,
thus providing the energy to these extended sources 
(e.g., Brunetti et al. 2001, Petrosian 2001). 

To date about 30 radio halos are known 
(e.g., Giovannini \& Feretti 2000, Bacchi et al. 2003,
Govoni et al. 2001, Venturi et al. 2007, 2008, 
Giovannini et al. in preparation).
Because of their extremely low surface brightness
and large angular extent ($>$10$'$ at a redshift z$\le$0.1)
radio halos are most appropriately studied at low
spatial resolution. Several radio halos were detected
by Giovannini et al. (1999)  in 
the NRAO VLA Sky Survey (NVSS; Condon et al. 1998) and by 
Kempner \& Sarazin (2001) in the Westerbork Northern Sky Survey
(WENSS; Rengelink et al. 1997), 
where the relatively large beam of these
surveys provide of the necessary sensitivity to large-scale 
emission in identifying these elusive sources.

A major merger event is expected to disrupt cooling cores
and create disturbances that are readily visible in
an X-ray image of the cluster. Therefore,
the merger scenario predicts the absence of large-scale radio
halos in symmetric cooling-core clusters.
However, a few cooling-core clusters exhibit signs of
diffuse synchrotron emission
that extends far from the dominant radio
galaxy at the cluster center, forming what is referred to as a mini-halo.
These diffuse radio sources are extended
on a moderate scale (typically $\simeq$ 500 kpc) and,
in common with large-scale halos, have a steep spectrum and a very
low surface brightness.

Because of a combination of small angular size and the strong radio 
emission of the central radio galaxy, 
the detection of a mini-halo requires data of a much higher dynamic range 
and resolution than those in available surveys,
and this complicates their detection.
As a consequence, our current observational knowledge on
mini-halos is limited to only a handful of clusters
(e.g., Perseus: Burns et al. 1992; A2390: Bacchi et al. 2003;
RXJ1347.5-1145: Gitti et al. 2007), and their origin and
physical properties are still poorly known.
The study of radio emission from the center of cooling-core 
clusters is of large importance not only in understanding the
feedback mechanism involved in the energy transfer between the
AGN and the ambient medium (e.g., McNamara \& Nulsen 2007) but also
in the formation process of the non-thermal mini-halos.
The energy released by the central AGN may also play a role
in the formation of these extended structures (e.g. Fujita et al. 2007).

On the other hand, the radiative lifetime of the relativistic electrons in
mini-halos is of the order of $\simeq$10$^7$$-$10$^8$ yrs, 
much shorter than the
time necessary for them to diffuse from the central radio galaxy
to the mini-halo periphery.
Thus, relativistic electrons must be reaccelerated and/or injected in-situ
with high efficiency in mini-halos.
Gitti et al. (2002) suggested that the mini-halo emission
is due to a relic population of relativistic electrons reaccelerated by
MHD turbulence via Fermi-like processes, the necessary
energetics being supplied by the cooling flow.
In support of mini-halo emission being triggered by the central cooling flow,
Gitti et al. (2004) found a trend between 
the radio power of mini-halos and the cooling flow power. 
Although mini-halos are usually found in
cooling-core clusters with no evidence of major mergers,
signatures of minor-merging activities
and gas-sloshing mechanisms in clusters containing mini-halos
(e.g., Gitti et al. 2007, Mazzotta \& Giacintucci 2008) have been revealed,
suggesting that turbulence related to
minor mergers could also play a role in the electron acceleration.

Alternatively, Pfrommer \& En{\ss}lin (2004) proposed that relativistic 
electrons in mini-halos
are of secondary origin and thus continuously produced by the interaction of
cosmic ray protons with the ambient, thermal protons.

Cassano et al. (2008) found that the synchrotron emissivity
(energy per unit volume, per unit time, per unit frequency) 
of mini-halos is about a factor of 50 higher than that of radio halos.
In the framework of the particle re-acceleration scenario,
they suggested that, an extra amount of relativistic electrons would be
necessary to explain the higher radio emissivity of mini-halos. These
electrons could be provided by the central radio galaxy or be of secondary origin.
 
To search for new extended diffuse radio emission in relaxed and 
cool-core galaxy clusters, we performed deep observations
of A1068, A1413, A1650, A1835, and A2029, carried out 
with the Very Large Array at 1.4 GHz, and complemented our data set
with a VLA archival observation of Ophiuchus.
Here, we present the new mini-halos that we identified in these data.
In Murgia et al. (submitted, hereafter Paper II), 
we quantitatively investigate the 
radio properties of these new sources and compare them with 
the radio properties of a statistically significant 
sample of mini-halos and halos 
already known in the literature, for which high quality VLA radio
 images at 1.4 GHz are available.

The radio observations and data reduction 
are described in Sect. 2. For each cluster, in Sect. 3, we 
investigate the possible presence of a central mini-halo.
In Sect. 4, we discuss the interplay between the mini-halos 
and the cluster X-ray emission. In Sect. 5, we analyze
a possible connection between the central cD galaxy 
and the surrounding mini-halo. Finally, our conclusions are 
presented in Sect. 6.

Throughout this paper, we assume a $\Lambda$CDM cosmology with
$H_0$ = 71 km s$^{-1}$Mpc$^{-1}$,
$\Omega_m$ = 0.27, and $\Omega_{\Lambda}$ = 0.73.

\section{VLA observations and data reduction}

To investigate the presence of diffuse, extended, radio emission
in relaxed systems, we analyzed new and archival VLA data of 
cooling-core clusters.
The list of the clusters is reported in Table 1, while the details of the
observations are described in Table 2.

\begin{table}
\caption{List of the relaxed clusters analyzed in this work.}
\begin{center}
\begin{tabular} {cccc} 
\hline
Cluster   & z &    kpc/$''$     & $D_L$\\
          &   &                 & Mpc  \\
\hline
A1068      &   0.1375    & 2.40 & 641.45  \\
A1413      &   0.1427    & 2.48 & 667.98  \\
A1650      &   0.0843    & 1.56 & 379.24  \\
A1835      &   0.2532    & 3.91 & 1267.62 \\
A2029      &   0.0765    & 1.43 & 342.24  \\
Ophiuchus  &   0.028     & 0.55 & 120.84  \\ 
\hline
\multicolumn{4}{l}{\scriptsize Col. 1: Cluster Name; Col. 2: Redshift;}\\
\multicolumn{4}{l}{\scriptsize Col. 3: Angular to linear conversion;}\\
\multicolumn{4}{l}{\scriptsize Col. 4: Luminosity distance.}\\
\end{tabular}
\label{tab1}
\end{center}
\end{table}

\begin{table*}
\caption{Details of the VLA observations.}
\begin{center}
\begin{tabular}{ccccccccc}
\hline 
\noalign{\smallskip}
Cluster & RA       & DEC       &  Frequency & Bandwidth & Config. & Obs. Time & Date & Note    \\
       &  J2000    & J2000     &  MHz     & MHz        &         & hours     &       &     \\
\noalign{\smallskip}
\hline
\noalign{\smallskip}
A1068 & 10 40 47.0     & $+$39 57 18.0   & 1465/1415 & 25 & C  & 3.4,1.0,5.7 & 2006 Oct 21,23,28 &  \\
A1413 & 11 55 19.0     & $+$23 24 30.0   & 1465/1415 & 25 & C  & 8.0         &  2006 Oct 23  & \\
A1650 & 12 58 41.0     & $-$01 45 25.0   & 1465/1415 & 25 & C  & 7.0          &  2006 Oct 22     & \\
A1835 & 14 01 02.0     & $+$02 51 30.0   & 1465/1385 & 50 & D  & 4.2          & 2003 March 10    & \\ 
A1835 & $''$           & $''$            & 1465/1415 & 25 & C  & 0.7,4.7      & 2006 Oct 23,28   & \\  
A1835 & 14 01 02.1     & $+$02 52 42.7   & 1465/1665 & 25 & A  & 2            & 1998 Apr 23      & archive (AT211)\\     
A2029 & 15 10 58.0   & $+$05 45 42.0   & 1465/1385 & 50     & D       & 4.7 & 2003 March 09      & \\
A2029 &  $''$        & $''$            & 1465/1415 & 50     & C       & 6.3 & 2006 Dec 26        & \\
A2029 & 15 10 56.1   & 05 44 42.6      & 1465/1515 & 25     & A       & 0.1 & 1993 Jan 14        &archive (AL252) \\
Ophiuchus & 17 12 31.9 & $-$23 20 32.6 & 1452/1502 & 25     & D       & 0.5 & 1990 Jan 25        &archive (AC261)\\    

\noalign{\smallskip}
\hline
\multicolumn{9}{l}{\scriptsize Col. 1: Cluster Name; Col. 2, Col. 3: Observation pointing 
(RAJ2000, DECJ2000); Col. 4: Observing frequency;}\\
\multicolumn{9}{l}{\scriptsize Col. 5: Bandwidth; Col. 6: VLA configuration; 
Col. 7: Total integration time on source; Col. 8: Observing dates;
Col. 9: Notes on archival observations.}\\
\end{tabular}
\label{data}
\end{center}
\end{table*}

We selected a small sample of 5 relaxed 
galaxy clusters: A1068, A1413, A1650, A1835, and A2029. 
They were selected on the basis of their extremely
regular cluster-scale X-ray morphology and lack of
evidence of a recent major merger.
All have been well observed by the Chandra satellite.
The X-ray data indicate that they all have cool,
dense cores (see references in sections on individual
clusters below). As often observed in these clusters
(e.g., Markevitch et al.\ 2003), all exhibit signs of the
sloshing of dense gas around their cD galaxy, the
presumed center of the cluster gravitational potential.
We also selected these cooling core clusters because
of their intermediate redshifts (z$>$0.05).
This selection criterion ensures that a mini-halo of 500 kpc in extent
would have an angular size smaller than 10$'$. Therefore,
there would be no significant missing flux for the shortest of
Very Large Array (VLA) baselines.
We observed A1068, A1413, and A1650 at 1.4 GHz with the VLA in 
C configuration, while A1835 and A2029 were observed 
both in C and D configurations.
These data provided an excellent combination of resolution 
and sensitivity in studying the large-scale cluster emission.

Calibration and imaging were performed with the
NRAO Astronomical Image Processing System (AIPS)
package. 
The data were calibrated in both phase and amplitude.
The phase calibration was completed using nearby secondary calibrators,
observed at intervals of $\sim$20 minutes. 
The flux-density scale was calibrated by observing 3C\,286. 
Brightness images were produced
following the standard procedures: Fourier-Transform, 
Clean, and Restore. Self-calibration was applied to remove residual
phase variations.

In C array, the 1$\sigma$ rms noise reached in our observations
range from 0.022 mJy/beam (in A2029) to 0.045 mJy/beam (in A1650),
while in D array the rms noise is of 0.025 mJy/beam in A1835
and 0.031 mJy/beam in A2029. 
The cluster A2029 also corresponds to the highest peak intensity
and in this case we reached a extraordinary high dynamic range.
The dynamic range, measured to be the ratio of the peak brightness to
the rms noise, was 10850:1 and 14450:1 in the C and D configuration 
respectively.

To isolate the mini-halo emission from the radio emission 
of the central cD galaxy, we retrieved high resolution archival observations 
at 1.4 GHz with the VLA in A configuration of A1825 and A2029 
(program AT211 and AL252 respectively).

While a wide-field X-ray image of the Ophiuchus cluster
from ROSAT PSPC suggests merging on a cluster-wide scale,
the Chandra X-ray image indicates that its cool core is 
insignificantly disturbed (apart from the usual sloshing,
Ascasibar \& Markevitch 2006), thus we decided to include
this interesting cluster in our sample.

We retrieved an archival observation of the Ophiuchus cluster
at 1.4 GHz with the VLA in D 
configuration (program AC261).
After precessing the original data to J2000
coordinates, the data were calibrated in phase and amplitude, 
and a cluster radio image was obtained following
standard procedures.
The phase calibration was completed using the nearby secondary calibrator
$1730-130$, observed at intervals of $\sim$10 minutes, 
while the flux-density scale 
was calibrated by observing 3C\,286. 
In this observation, we achieved a 1$\sigma$ rms noise level of 0.1 mJy/beam.
At the distance of the Ophiuchus cluster, a mini-halo of 500 kpc in extent
would have an angular size of about 15$'$. 
This value is close to the maximum angular size observable at 1.4 GHz 
by the VLA in D configuration.

\section{Results} 

For each cluster, we report the results obtained in
this campaign of observations.
We find that in the central regions of A1835, A2029, and Ophiuchus, 
the dominant radio galaxy is surrounded by a diffuse low-brightness 
radio emission that we identify as a mini-halo.
Low-significance indication of diffuse emission are present  
in A1068 and A1413, although their unambiguous classification as mini-halos  
would require further investigation.
Finally, we detect no diffuse emission in A1650.
We indicate the radio flux densities of both the mini-halos and the
central point sources in A1835, A2029, and Ophiuchus, calculated
by using the fit procedure described in Paper II,
for more details of the density flux calculation, see that paper.

\subsection{Abell 1068} 

A1068 is a relaxed cluster that has a peaked X-ray
surface brightness profile, a declining temperature gradient, and
an increasing gas metallicity toward the cluster center 
(Wise et al. 2004, McNamara et al. 2004), 
exhibiting many features commonly seen in other cooling-core clusters.

\begin{figure*}
\begin{center}
\includegraphics[width=15cm]{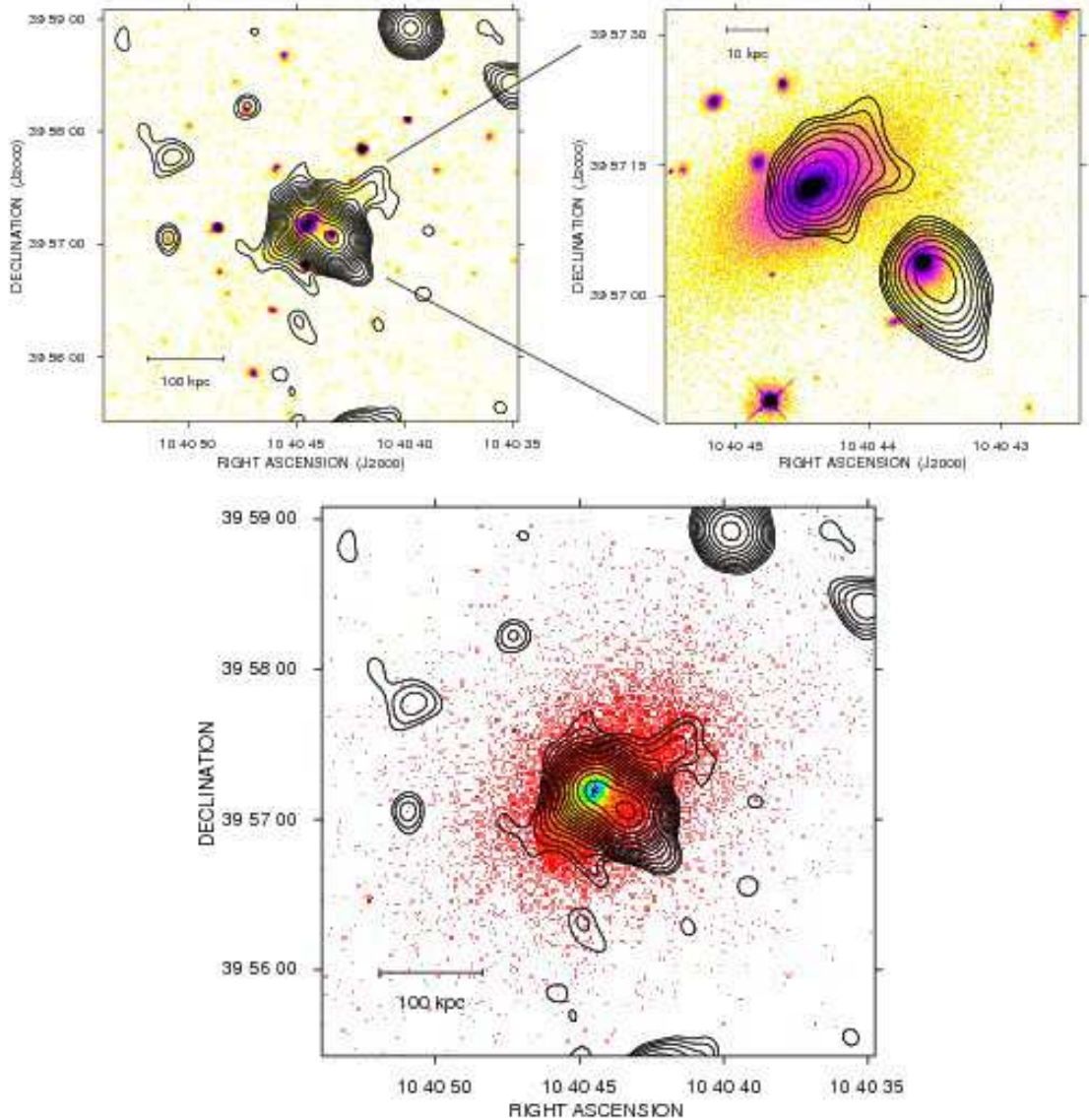}
\caption{Top, left: total intensity radio contours of A1068 at 1.4 GHz with a
FWHM of 15$''\times 15''$. 
The first contour level is drawn at 0.08 mJy/beam
and the others are spaced by a factor $\sqrt{2}$. The sensitivity 
(1$\sigma$) is 0.027 mJy/beam. 
The contours of the radio intensity are 
overlaid on the optical POSS2 image.
Top, right: zoom of total intensity radio contours in the center of A1068 
at 1.4 GHz with a FWHM of 5.4$''\times 5.4''$ taken from the FIRST survey. 
The first contour level is drawn at 0.5 mJy/beam
and the rest are spaced by a factor $\sqrt{2}$. The sensitivity 
of the FIRST survey (1$\sigma$) is 0.15 mJy/beam. 
The contours of the radio intensity are overlaid on the optical image
taken from the HST (F606W filter) archive.
Bottom: total intensity radio contours of A1068 at 1.4 GHz with a
FWHM of 15$''\times 15''$ overlaid on the Chandra X-ray image in the 0.5-4 keV
band.}
\label{A1068}
\end{center}
\end{figure*}

Figure 1 shows the central cluster radio emission
compared with optical and X-ray images.
In the top left panel, we present the optical cluster image
with a radio contour plot overlaid.
The total intensity radio contours are from the 1.4 GHz image 
with a resolution of 
15$''$, while the optical image was taken from the
POSS2 red plate in the Optical Digitalized Sky Survey  
\footnote{See http://archive.eso.org/dss/dss}.
The field of view of this image is about 3.5$'$.
This panel suggests that  diffuse emission
on a scale larger than 100 kpc may be present around the central 
radio emission. The central radio emission exhibits two peaks.
The first coincides with the cluster cD galaxy, while the second
is located toward the south-west at a distance of about 15$''$ 
and is associated with another bright cluster galaxy.
The separation of these two radio galaxies is shown in the top right 
panel of Fig. 1, where the higher resolution ($\simeq$5$''$) 
FIRST image (Becker, White \& Helfand 1995) is overlaid on the HST image. 
The radio emission associated with the
cD galaxy has a flux density of 8.5$\pm$0.6 mJy, which corresponds to a
power $L_{1.4GHz}=4.2\times10^{23}$W/Hz. It is marginally resolved 
and extends to the north-west.
The south-western source shows a head-tail morphology and has a 
flux density of  10.4$\pm$0.6 mJy.

To highlight the possible diffuse large-scale emission
associated with the intra-cluster medium,
we subtracted the flux density of the two central sources, calculated in the 
FIRST survey from our lower resolution image.
The total radio emission estimated in the cluster central emission from
our map was 22.3$\pm$0.7 mJy. Therefore, 3.4$\pm$1.1 mJy, which corresponds to a
power $L_{1.4GHz}=1.7\times10^{23}$W/Hz, appears to be associated 
with low diffusion emission.
However, we note that the FIRST subtraction may be uncertain.
A possible variation in the
core-component density flux and/or any absolute calibration 
error between the FIRST and this dataset could cause in an under or
over subtraction of flux, and therefore the residual flux
associated with low diffusion emission must be interpreted with caution.

In the bottom panel, our radio image is overlaid on the 
Chandra X-ray emission of the cluster in the $0.5-4$ keV band.
The X-ray emission appears elongated in the north-west south-east
direction. The low surface brightness radio emission that is
possibly associated with cluster diffuse emission appears to
have the same orientation as both the cluster X-ray emission 
and the central cD galaxy.

In conclusion, a hint of diffuse emission has been 
detected in our data around the dominant cluster galaxy in A1068. 
However, given the uncertainty related to
comparing datasets, this result requires further confirmation.

We note that in the field of view of this observation, 
we detected a giant radio galaxy that is not related to A1068, 
whose properties are shown in Appendix A.

\subsection{Abell 1413}

\begin{figure*}
\begin{center}
\includegraphics[width=15cm]{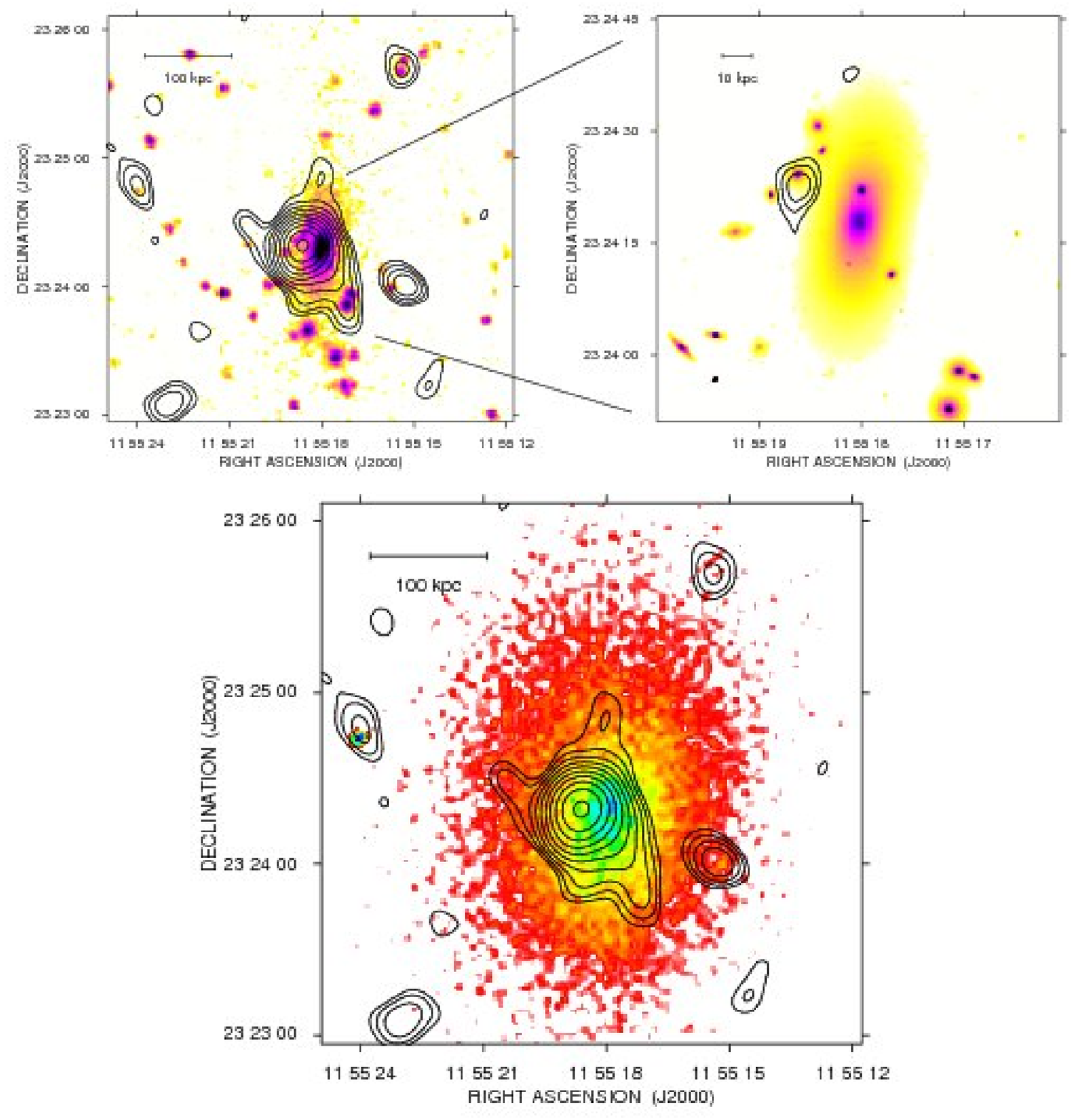}
\caption{
Top, left: total intensity radio contours of A1413 at 1.4 GHz with a
FWHM of 15$''\times 15''$.
The first contour level is drawn at 0.1 mJy/beam
and the rest are spaced by a factor $\sqrt{2}$.
The sensitivity (1$\sigma$) is 0.035 mJy/beam.
The contours of the radio intensity are overlaid on the optical 
POSS2 image.
Top, right: zoom of total intensity radio contours in the center of A1413 
at 1.4 GHz with a FWHM of 5.4$''\times 5.4''$ taken from the FIRST survey. 
The first contour level is drawn at 0.5 mJy/beam
and the rest are spaced by a factor $\sqrt{2}$. The sensitivity 
of the FIRST survey (1$\sigma$) is 0.15 mJy/beam.
The contours of the radio intensity are overlaid on the optical image
taken from the HST (F606W filter) archive.
Bottom: total intensity radio contours of A1413 at 1.4 GHz with a
FWHM of 15$''\times 15''$ overlaid on the Chandra X-ray image in the 0.5-4 keV
band.
}
\label{A1413}
\end{center}
\end{figure*}

In the top left panel of Fig. 2, our deep radio observation
at 1.4 GHz with a 
resolution of 15$''$ is overlaid on the POSS2 red plate image.
The field of view of the image is about 3$'$. 
A diffuse low-surface-brightness emission of about 1.5$'$
($\simeq 220$ kpc) in size was detected in close proximity to
the cluster center.
However, the radio-optical overlay clearly indicates that the peak
of the central radio emission is offset to the east with respect 
to the central cD galaxy. 

To investigate in detail the central radio
emission of the cluster, in the top right panel of 
Fig. 2 we show a high resolution zoom of this region.
In this panel, the image at 1.4 GHz 
taken from the FIRST survey with a resolution of $\simeq$5$''$ is 
overlaid on the HST image. 
The FIRST survey confirms that any discrete radio emission 
is associated with the central
cD. We estimated
a 3$\sigma$ flux limit to a point-source radio emission 
associated with the central cD galaxy of 0.45 mJy, which corresponds to 
a power $L_{1.4GHz}<2.4\times10^{22}$W/Hz.
A faint (flux density of 2.9$\pm$0.7 mJy), unresolved, 
radio source is detected that coincides with another galaxy 
to the east.

To estimate the flux density of diffuse emission on large scales,
we subtracted the flux density of the faint radio source detected in 
the FIRST survey from that measured in our lower resolution image.
The total radio emission estimated in the cluster central emission from
our map was 4.8$\pm$0.2 mJy. 
Therefore 1.9$\pm$0.7 mJy, which corresponds to a
power $L_{1.4GHz}=1.0\times10^{23}$W/Hz, is indicative
of low diffusion emission, but the low significance ($<3\sigma$)
require further observations. As mentioned in the case of A1068, 
the FIRST flux density subtraction must also be
considered with caution.

In the X-ray band, A1413 has been observed by XMM-Newton (Pratt \&
Arnaud 2002) and Chandra (Baldi et al. 2007, Vikhlinin et al 2005).
It exhibits a regular morphology and no evidence of recent merging. 
In the bottom panel of Fig. 2, the previous radio image is overlaid on the 
Chandra X-ray emission of the cluster in the $0.5-4$ keV band.
The orientation of the X-ray emission is well aligned with that of the
central cD galaxy, while the diffuse radio emission 
is offset from the X-ray peak and seems
elongated in a slightly different direction.
However this could be due to the confusion effect of the nearby unrelated
discrete source visible in the FIRST image. As in the case of A1068, 
further observations are needed to
confirm the presence of a mini-halo and the possible discrepancy between the
radio and the X-ray peak.

A peculiarity of the extended emission in A1413 is  
that the cD galaxy, located in the middle of this putative mini-halo, 
does not contain a compact radio source, 
at least at the FIRST sensitivity level.
Another cluster with a similar characteristic is A2142 (Giovannini \&
Feretti 2000). The properties of these clusters may suggest the presence 
of ``relic''
mini-halos in which the central cD galaxy is switched-off, while the diffuse
mini-halo continues to emit. 
Observations at lower frequency might help in understanding this case.

\subsection{Abell 1650}

\begin{figure*}
\begin{center}
\includegraphics[width=15cm]{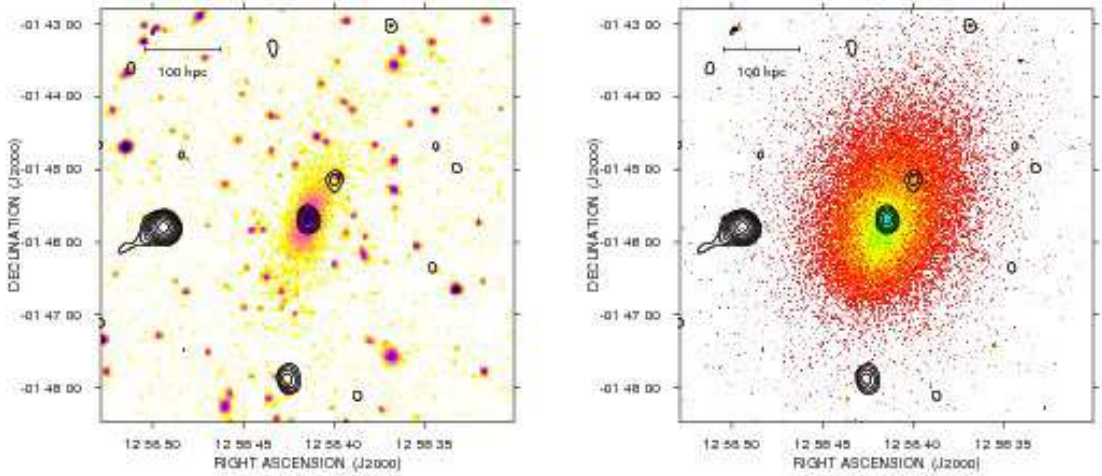}
\caption{Left: total intensity radio contours of A1650 at 1.4 GHz with a
FWHM of 16$''\times 16''$.
The first contour level is drawn at 0.13 mJy/beam
and the rest are spaced by a factor $\sqrt{2}$.
The sensitivity (1$\sigma$) is 0.045 mJy/beam.
The contours of the radio intensity are overlaid on the optical 
POSS2 image.
Right: total intensity radio contours of A1650 overlaid 
on the Chandra X-ray image in the 0.5-4 keV band.
}
\label{A1650}
\end{center}
\end{figure*}

Studies of this cluster in X-ray
(Takahashi \& Yamashita 2003, Donahue et al. 2005)
pointed out that there is no clear evidence of substructures,
suggesting that the cluster has not experienced a major
merger in the recent past, and is in a relaxed state.
However, while the central cooling time of this cluster is shorter than the 
Hubble time and it has a strong metallicity gradient,
it does not have a significant temperature gradient close to its center.

A1650 contains a single optically luminous central cD galaxy that is
known to be radio quiet (Burns 1990). The FIRST survey
does not detect any radio emission coincident with this 
dominant cluster galaxy.

We imaged this cluster at 1.4 GHz, reaching a sensitivity
at 1$\sigma$ of 0.045 mJy/beam. 
In Fig. 3, the central cluster radio emission is overlaid 
on both the POSS2 and the Chandra image.  
We note that our deep image reveals the presence 
of a faint, unresolved, radio source in coincidence with the cD galaxy.
Its flux density is 0.44$\pm$0.09 mJy, which corresponds to a
power $L_{1.4GHz}=7.6\times10^{21}$W/Hz.
However, at this sensitivity level, 
this cluster does not exhibit any signs of mini-halo.
We estimated a 3$\sigma$ flux limit of 1.35 mJy for a mini-halo extended 
300 kpc in size, which corresponds to a power of
$L_{1.4GHz}<2.3\times10^{22}$W/Hz.   

At the periphery of the A1650 cluster, we note that 
we detected an unusually round radio galaxy, whose properties 
are shown in Appendix B.

\subsection{Abell 1835}

The X-ray emission of A1835 has been studied with {\it XMM} 
(Peterson et al. 2001, Majerowicz et al. 2002) and {\it Chandra} 
(Schmidt et al. 2001, Markevitch 2002), and both
sets of observations provided evidence that the cluster 
has a relatively cool ($3-4$ keV) inner core surrounded by a hotter
($8-9$ keV) outer envelope.
Overall, the X-ray image displays a relaxed morphology, although 
substructures have been detected at the cluster center.

\begin{figure*}
\begin{center}
\includegraphics[width=15cm]{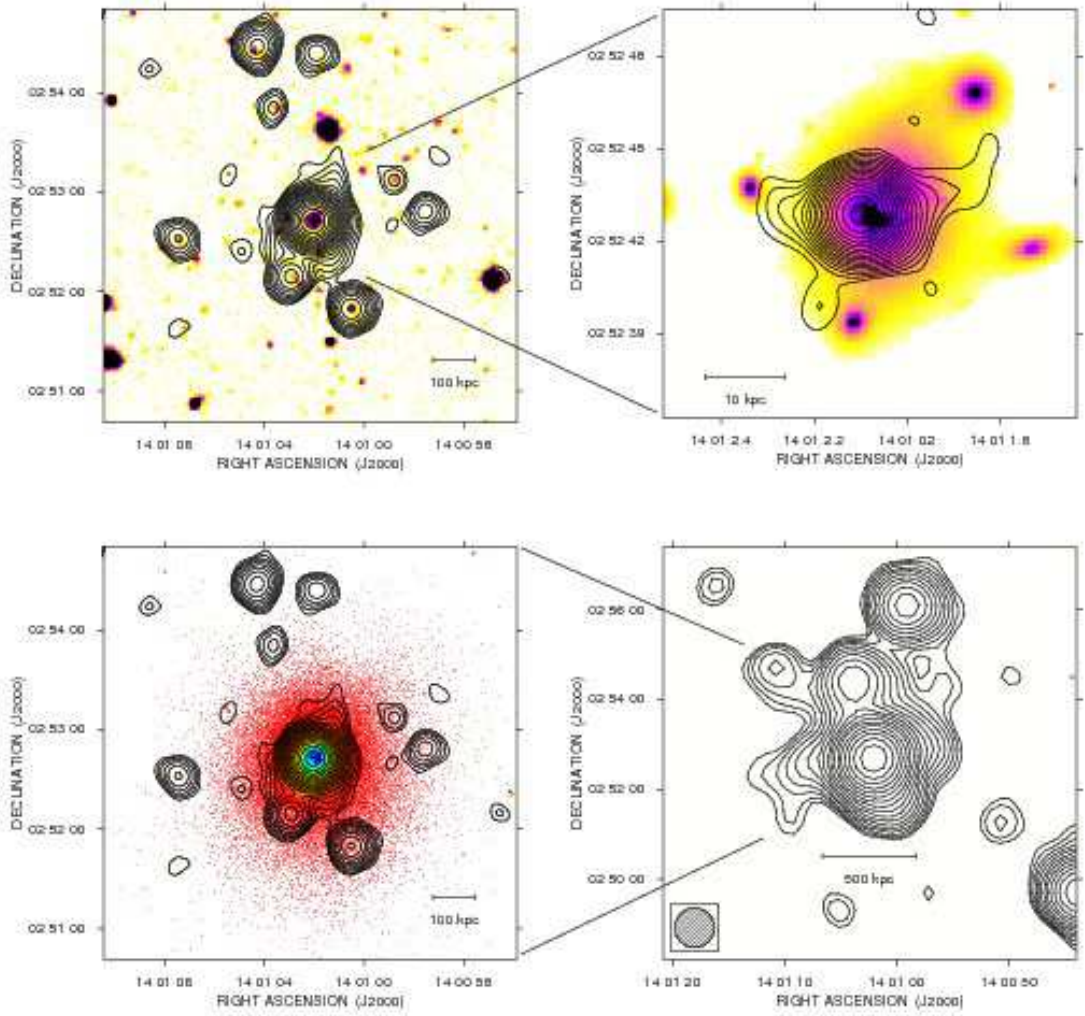}
\caption{Left: total intensity radio contours of A1835 at 1.4 GHz with a
FWHM of 16$''\times 16''$. 
The first contour level is drawn at 0.1 mJy/beam
and the rest are spaced by a factor $\sqrt{2}$.
The sensitivity (1$\sigma$) is 0.036 mJy/beam.
The contours of the radio intensity are overlaid on the optical 
POSS2 image (top) and on the Chandra X-ray image in the 0.5-4 keV band
(bottom).
Right, top: zoom of total intensity radio contours in the center of A1835 
at 1465 MHz with a FWHM of 1.36$''\times 1.30''$ ($PA=2.5^0$) taken from a
VLA archival data set (AT211). 
The first contour level is drawn at 0.12 mJy/beam
and the rest increase by a factor $\sqrt{2}$. The sensitivity 
of the high resolution image (1$\sigma$) is 0.037 mJy/beam.
The contours of the radio intensity are overlaid on the optical image
taken from the HST (F702W filter) archive.
Right, bottom: larger field of view of the
total intensity radio contours of A1835 at 1.4 GHz with a
FWHM of 53$''\times 53''$.
The first contour level is drawn at 0.1 mJy/beam
and the rest are spaced by a factor $\sqrt{2}$.
The sensitivity (1$\sigma$) is 0.025 mJy/beam.
}
\label{A1835}
\end{center}
\end{figure*}

In Fig. 4, we display the central cluster radio emission, obtained 
with the VLA in C configuration, superimposed on optical and X-ray images.
In the top left panel, we present the optical cluster image
with a radio contour plot overlaid. The total intensity radio contours
are at 1.4 GHz with a resolution of  16$''$, while the optical
image was extracted from the  POSS2 red plate in the
Optical Digitalized Sky Survey.
The field of view of this image is about 4$'$.
Our image clearly displays evidence of extended emission around the
central radio core.

The peak of the central radio emission is coincident
with the cluster cD galaxy.
The emission as detected at higher resolution is shown in
the top right panel of Fig. 4, in which we overlay
the high resolution radio image from the VLA
archive (project AT211) on the HST image.
The sensitivity of the image is 
0.037 mJy/beam and the resolution is $1.36''\times1.30''$.

In the bottom left panel, our 16$''$ resolution radio image is overlaid on the 
Chandra X-ray emission of the cluster in the $0.5-4$ keV band.
The diffuse radio emission 
appears extended to several
hundreds of kiloparsecs and appears well mixed with 
the thermal plasma.  
The round morphology of the diffuse emission  
also coincides well with the morphology of the thermal gas.

The bottom right panel displays a large field of view of the
total intensity radio contours of A1835 at 1.4 GHz with a
FWHM of 53$''\times 53''$. This image was obtained with the VLA in
D configuration.
Despite the presence of several radio galaxies located
close to the cluster center, a new mini-halo in A1835 is clearly 
evident at this lower resolution.

The mini-halo flux density is $6.0\pm 0.8$ mJy,  
which corresponds to a power $L_{1.4GHz}=1.2\times10^{24}$W/Hz.
The central radio source has a flux density of $32.2\pm 0.8$, 
which corresponds to a
power $L_{1.4GHz}=6.2\times10^{24}$W/Hz, which agrees 
with  B{\^i}rzan et al. (2008).

\subsection{Abell 2029}

The nearby well-studied cluster of galaxies A2029, is one of 
the most optically detected regular rich clusters known
and is dominated by a central ultra-luminous 
cD galaxy (Dressler 1978, Uson et al. 1991).

The hot cluster A2029 has been extensively studied in X-rays 
(e.g., Buote \& Canizares 1996, Sarazin et al.\ 1998, 
Lewis et al.\ 2002, Clarke et al.\ 2004, Vikhlinin et al.\ 2005, 
Bourdin \& Mazzotta 2008). On scales $r>100-200$ kpc, it is one of the most
regular and relaxed clusters known (e.g., Buote \& Tsai
1996).  However, in the image of the cool dense core, Clarke
et al. (2004) observed a subtle spiral structure. Along with
several sharp brightness edges, it indicates ongoing gas
sloshing that is most likely indicative of past subcluster
infall episodes (Ascasibar \& Markevitch 2006). As in many
cool-core clusters with central radio sources, some
small-scale X-ray structure is also apparently connected with
the central radio galaxy (Clarke et al. 2004)

In the top left panel of Fig. 5, our deep radio observation
at 1.4 GHz with a 
resolution of 16$''$, obtained with the VLA in C configuration,
is overlaid on the POSS2 red plate image.
The field of view of the image is about 7$'$. 
A large-scale diffuse emission,
most likely a mini-halo, is located around, but has 
its centroid offset from
the central PKS1509+59 source, which is coincident with the
cluster cD galaxy.
The mini-halo is far more extended than the
central radio galaxy and appears to have a filamentary
morphology slightly elongated in a north-east to south-west direction.
The relatively high resolution of our new image 
ensures that the detected diffuse emission is real and
not due to a blend of discrete sources.
A sign of a diffuse, extended emission in this cluster 
is also found by Markovi{\'c} et al. (2003) in a VLA observation at 74 MHz.

To distinguish more effectively between the mini-halo emission and
the radio emission of the central cD galaxy, in the top right panel of 
Fig. 5, we show a high resolution zoom of the cluster central region.
In this panel, the image at 1.4 GHz with a resolution of
$1.62'' \times 1.35''$ taken from an archive data set (AL252) is 
overlaid on the HST image.
 As shown in this panel, 
the radio source PKS1509+59 associated with the central cD galaxy
has a distorted morphology and two oppositely directed jets.
The source has also a very steep integrated spectrum
and a high rotation measure as discussed in detail by
Taylor et al. (1994).

In the bottom left panel of Fig. 5, the previous radio image 
is overlaid on the 
Chandra X-ray emission of the cluster in the $0.5-4$ keV band.
We retrieved archival X-ray data for comparison of the 
gas distribution and the cluster radio emission.
The mini-halo appears elongated
in the same direction as both the  
cluster X-ray emission and the central cD galaxy.

The bottom right panel shows the
total intensity radio contours of A2029 at 1.4 GHz for a larger field of view
with a FWHM of 53$''\times 53''$. This image was obtained with the
VLA in D configuration. 
The mini-halo in A2029 is even clearer in this
lower resolution image.

The mini-halo flux density is $18.8\pm 1.3$ mJy, which corresponds 
to a power $L_{1.4GHz}=2.6\times10^{23}$W/Hz.
While the central radio source has a flux density of $480.0\pm 19.5 mJy$,  
which corresponds to a power $L_{1.4GHz}=6.7\times10^{24}$W/Hz.  

\begin{figure*}
\begin{center}
\includegraphics[width=15cm]{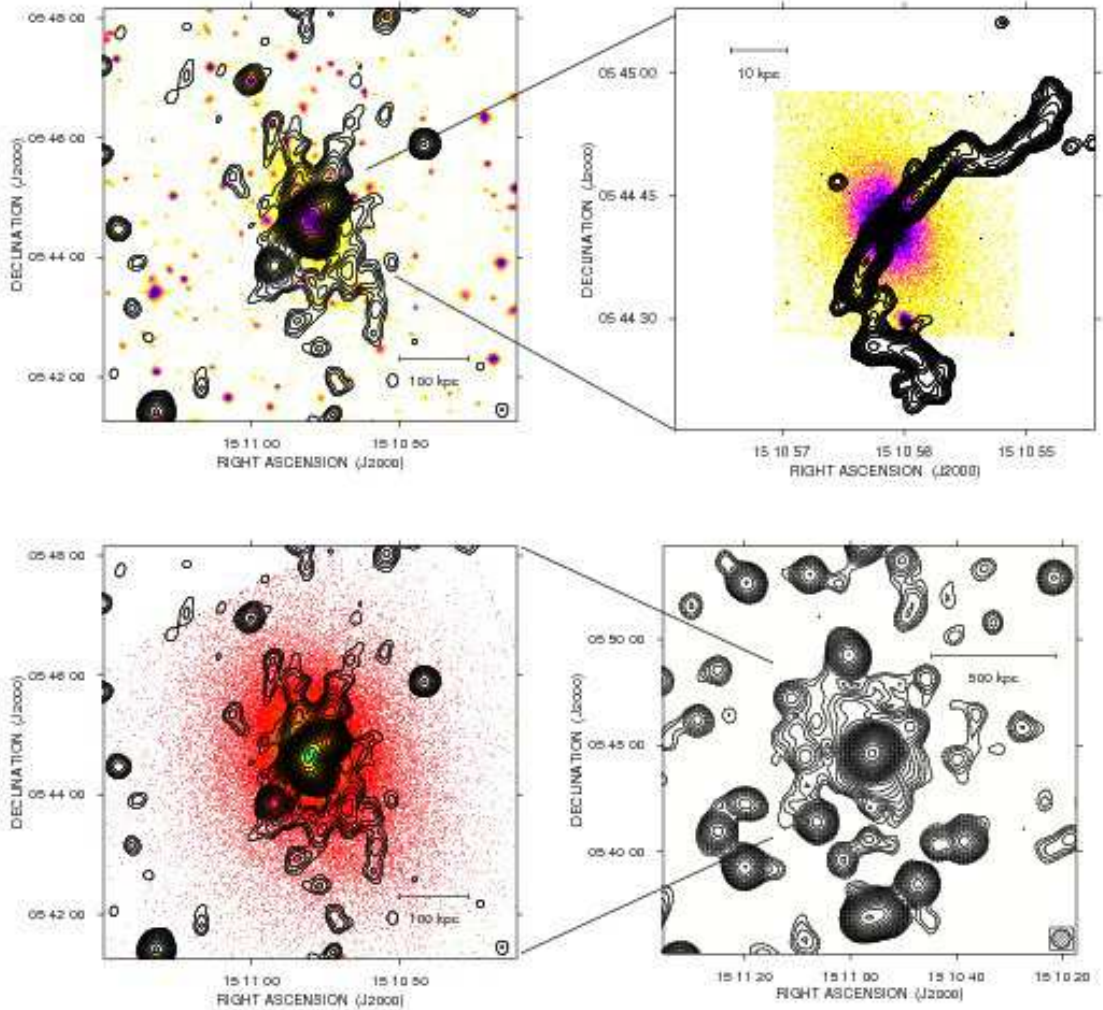}
\caption{
Left: total intensity radio contours of A2029 at 1.4 GHz with a
FWHM of 16$''\times 16''$. 
The first contour level is drawn at 0.07 mJy/beam
and the rest are spaced by a factor $\sqrt{2}$.
The sensitivity (1$\sigma$) is 0.022 mJy/beam.
The contours of the radio intensity are overlaid on the optical 
POSS2 image (top) and on the Chandra X-ray image in the 0.5-4 keV band
(bottom).
Right, top: zoom of the total intensity radio contours in the center of A2029 
at 1.4 GHz with a FWHM of 1.62$''\times 1.35''$ ($PA=-48.8^0$) taken from a
VLA archive data set (AL252). 
The first contour level is drawn at 0.25 mJy/beam
and the rest are spaced by a factor $\sqrt{2}$. The sensitivity 
of the high resolution image (1$\sigma$) is 0.083 mJy/beam.
The contours of the radio intensity are overlaid 
on the optical image taken from the HST (F450W filter) archive.
Right, bottom: larger field of view of the
total intensity radio contours of A2029 at 1.4 GHz with a
FWHM of 53$''\times 53''$.
The first contour level is drawn at 0.1 mJy/beam
and the rest are spaced by a factor $\sqrt{2}$.
The sensitivity (1$\sigma$) is 0.031 mJy/beam.
}
\label{A2029}
\end{center}
\end{figure*}

\subsection{Ophiuchus}

Ophiuchus is a nearby, rich cluster located
12 deg from the Galactic center.
In X-rays, it is one of the hottest clusters known.

\begin{figure*}
\begin{center}
\includegraphics[width=15cm]{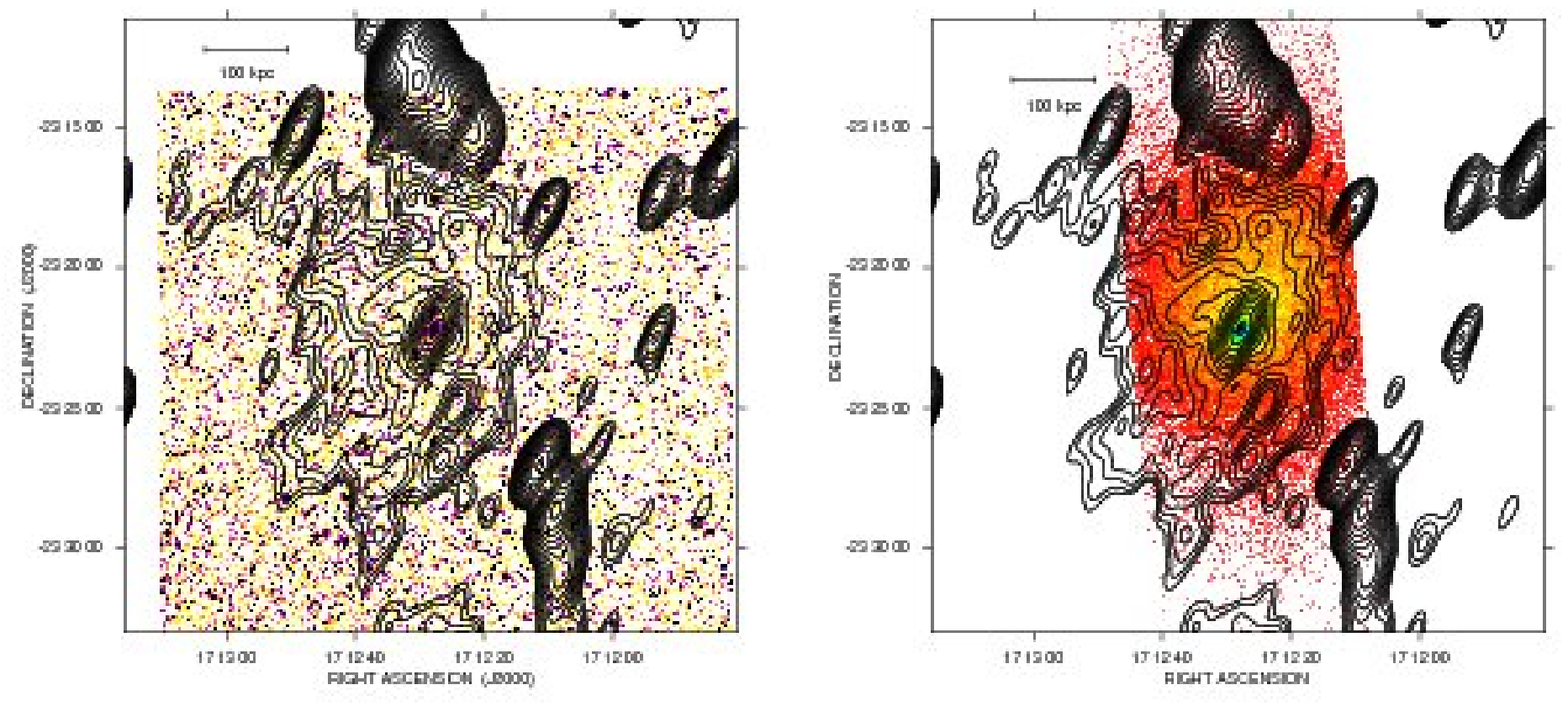}
\caption{Left: total intensity radio contours of Ophiuchus at 1.4 GHz 
with a FWHM of 91.4$''\times 40.4''$, $PA=-24.4^0$.
The first contour level is drawn at 0.3 mJy/beam
and the rest are spaced by a factor $\sqrt{2}$.
The sensitivity (1$\sigma$) is 0.1 mJy/beam.
The contours of the radio intensity are overlaid on the optical 
POSS2 image.
Right: total intensity radio contours of Ophiuchus overlaid 
on the Chandra X-ray image in the 0.5-4 keV band.
}
\end{center}
\label{Ophi}
\end{figure*}

Suzaku data (Fujita et al. 2008) and the archival Chandra
data indicate that this hot cluster has a cool, dense core.
The archival wide-field ROSAT PSPC image suggests an ongoing
unequal-mass merger, but the Chandra image indicates that
the core is still largely undisturbed, although it exhibits
the prototypical cold fronts caused by gas sloshing
(Ascasibar \& Markevitch 2006). 
Eckert et al. (2008) detected, with INTEGRAL,
X-ray emission at high energies 
in excess of a thermal plasma spectrum with the cluster's mean temperature. 
This emission may be of non-thermal origin, caused by for example,
by Compton scattering of relativistic electrons by 
the cosmic microwave background
radiation (see e.g., Rephaeli et al. 2008, Petrosian et al. 2008, and references
therein for reviews).
It would be particularly interesting to investigate the presence
of relativistic electrons in the intergalactic 
medium of this cluster because by assuming that the 
synchrotron emission and the hard X-ray excess are cospatial and 
produced by the same population of relativistic electrons, 
their detection would allow the determination of the 
cluster magnetic field.

We analyzed a VLA archive observation at 1.4 GHz in D configuration. 
In the left panel of Fig. 6, the radio observation
at 1.4 GHz with a 
resolution of $91.4'' \times 40.4''$ is overlaid on 
the POSS2 red plate image.
The asymmetric beam is due to the far southern declination of the cluster.
The field of view of the image is about 20$'$. 
It is evident that a large-scale low surface brightness diffuse emission,
which is probably a mini-halo, is located around the central cD galaxy.
We note that we may have missed some extended flux
due to the large angular extension of 
the mini-halo and/or because of the poor uv coverage in the short VLA 
observation (only 0.5 hours).

The mini-halo flux density, calculated on the basis of the
fit analysis described in Paper II, is $106.4 \pm 10.4$ mJy,  
which corresponds to a power $L_{1.4GHz}=1.9\times10^{23}$W/Hz.
The same analysis was used to estimate the radio flux density 
of the central cD galaxy, since no high resolution data for this cluster is available
in the VLA archive. We obtained a flux density of $29\pm 2$ mJy,  
which corresponds to a power $L_{1.4GHz}=5.1\times10^{22}$W/Hz.

In the right panel of Fig. 6, our radio image 
is overlaid on the Chandra X-ray emission of the cluster 
in the $0.5-4$ keV band.
The peak of the X-ray emission is coincident with the position
of the cD galaxy. 
Although the Chandra detector does not cover the entire cluster
X-ray extension, a connection between the mini-halo 
and the X-ray emission is evident, which is better investigated in the 
next section.

\section{Comparison between mini-halos and the X-ray emitting gas}

By superimposing the radio and X-ray data presented in previous sections,
we emphasized the similarity between the radio mini-halo
emission and the cluster X-ray morphologies of A2029, A1835, and Ophiuchus.
Because of the large angular extension of Ophiuchus, it is 
possible to perform a quantitative comparison between the radio 
and X-ray brightness.
Here we compare the Chandra count rate image in the $0.5-4$ keV band, 
with the VLA radio image, corrected for the primary beam attenuation.

We first constructed a square grid covering a region
containing both the radio mini-halo and the X-ray emission
from which we excluded areas containing discrete radio sources.
The grid cells size was chosen to be as large as the radio 
beam ($\simeq$ 90$''$).
In the statistical analysis, all the discrete sources were 
excluded by masking them out. All pixels lying in blanked areas 
or close to the edges of the chips were also excluded from the statistics. 
A scatter plot of the radio versus X-ray brightness 
comparisons is presented in top panel of Fig. 7. Each point represents 
the mean brightness in each cell of the grid, while the error bars
indicate the corresponding statistical error.
The close similarity between radio and X-ray structures in Ophiuchus
is demonstrated by the correlation between these two parameters.
We fitted the data with a power law relation of the type:
$I_{1.4 GHz}\propto I_X^{1.51\pm 0.04}$.
We then present the same correlation this time obtained by azimuthally 
averaging the radio and the X-ray emission in annuli (see middle panel
 of Fig.\,7). The data are fitted by a power law relation of the type
$I_{1.4 GHz}\propto I_X^{1.6\pm 0.1}$, which is consistent within the
errors with the scatter plot analysis.
The correlation indicates that the radial decline in the 
non-thermal radio component is slightly steeper than that of the thermal 
X-ray component, as can be seen in the bottom panel of Fig. 7.

This trend may be explained by secondary models of mini-halos, 
where a super-linear power law relation 
is expected (Dolag \& En{\ss}lin 2000, Pfrommer et al. 2008).
However, this result must be viewed with uncertainty, because 
as previously mentioned,
we may miss part of the radio emission on the largest angular scales,
which would reduce the slope of that relation.
Moreover, especially in cooling-core clusters, the slope
of the radius versus the X-ray brightness may be affected by strongly
variable temperature across the cluster.

It is interesting to compare the energetics of the mini-halos
with that of the thermal X-ray emitting gas. It has
become clear that AGNs play a crucial role in reheating the gas in the
cores of clusters (e.g., McNamara \& Nulsen 2007).  
This decelerates the cooling and condensing of
thermal gas onto the cD galaxy at the center of the cluster. The
details of reheating are still misunderstood, since the volume
occupied by the radio jets and lobes is small compared to that of the
cool core.  The mini-halo, however, appears well-matched to the cluster
core in extent, even corresponding in elongation and surface brightness
in some cases (e.g., Ophiucus, A2029).  

Applying standard minimum energy arguments to the radio emission from
the mini-halos and assuming that $k/f=1$ 
(where $k$ is the ratio of the relativistic 
particle energy to that in electrons emitting synchrotron radiation, and
$f$ is the volume filling factor of magnetic field and relativistic
particles) we infer a typical equipartition 
magnetic field strength of 0.6 $\mu$G 
(but see Paper II for more detailed calculations), and a total energy of roughly
$6 \times 10^{57}$ ergs for a typical mini-halo of radius 150 kpc.  
If we adopt a timescale for resupply of $10^7$ years 
then we derive an energy dissipation of
2 x $10^{43}$ erg/s.  This is smaller by a factor 50 than the $10^{45}$
erg/s required to balance cooling in the X-rays (Fabian 1994).  
A possible solution is to increase the value of $k/f$.  
Dunn, Fabian \& Taylor (2005) measured a wide range of 
$k/f$ values in clusters with values for the mini-halos typically
$\sim$100, though they did not single out that class of objects.  
Since the total energy increases only by 
$(k/f)^{4/7}$, increasing $k/f$ by a factor of 100 only increases
the total energy by a factor $\sim$14, leaving a factor $\sim$3.5 
unexplained. Given the various approximations employed, this 
is fairly close to the energy required.
However, it raises the questions of (1) what process supplies energy to
the mini-halos? and (2) how do the mini-halos couple with the thermal
gas?  Even if the mini-halos are not involved in 
the heating of the cluster, they could be
 tracing where the heating 
is going on.  One possibility is that the mechanism that heats the cluster
could also deposit some energy into relativistic electrons that then 
radiate in the radio.

\begin{figure}[h]
\begin{center}
\includegraphics[height=18cm]{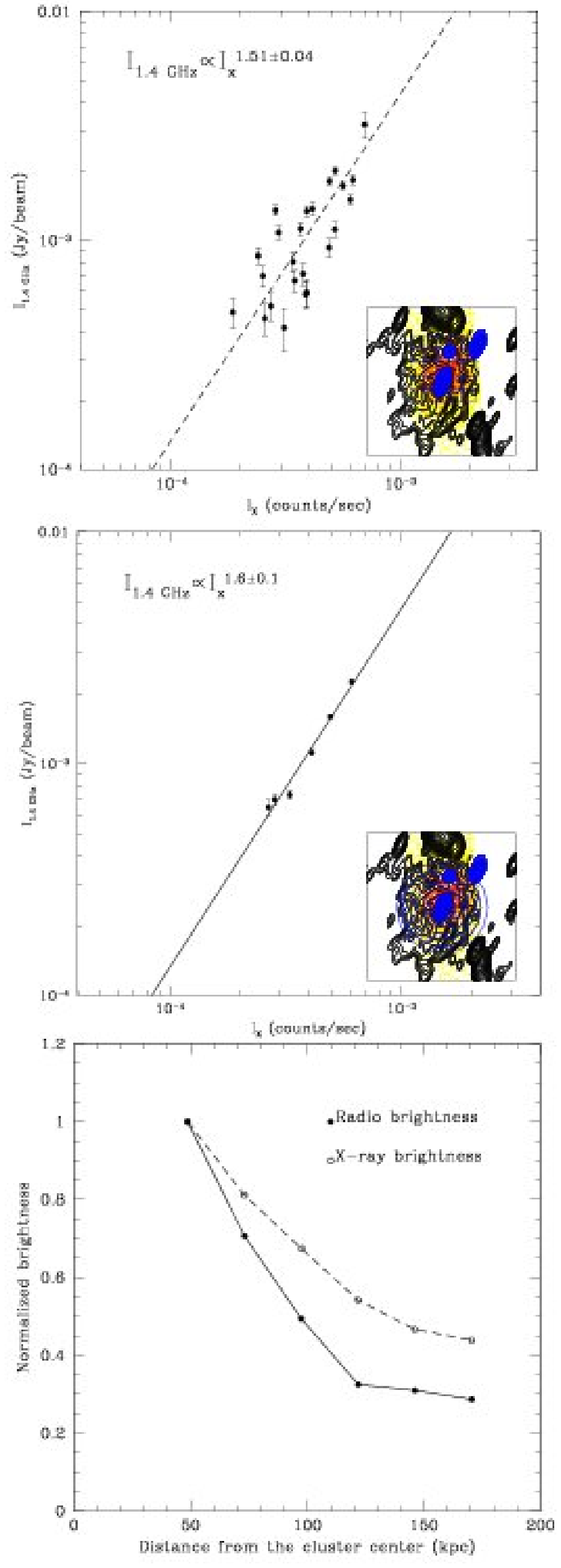}
\caption{Comparison of the radio and X-ray brightness in the
Ophiuchus galaxy cluster. The scattered plot of the radio intensity at 1.4 GHz versus the Chandra X-ray count rate is shown in the top panel. Each point represents the average in a rectangular region of $90''\times90''$ as shown in the inset. The middle panel presents the same correlation this time obtained by azimuthally averaging the radio and the X-ray emission in annuli as shown in the inset. The normalized X-ray and radio radial profiles are shown in the bottom panel. All pixels lying in blanked areas or close to the edges of the chips have been excluded from the statistics.}
\label{RadX}
\end{center}
\end{figure}

\section{Comparison between the mini-halos and the central cD radio galaxies}

As shown by Burns (1990), a large number of cooling core clusters
contain powerful radio sources associated with central cD galaxies.
Best et al. (2007) showed that brightest group and
cluster galaxies are more likely to host a radio-loud AGN
than other galaxies of the same stellar mass.

Since mini-halos are located at the center of cooling core clusters,
where a radio-loud AGN is in general present, a possible link
between the radio emission of the cDs and the mini-halos is interesting 
to investigate in the framework of the 
models attempting to explain the formation of mini-halos. 

In Fig. 8, we plot the radio power at 1.4 GHz of the mini-halos 
versus those of the central cD galaxies.
In addition to data for A1835, A2029 and Ophiuchus we plot data
for RXJ1347.5-1145 (Gitti et al. 2007), A2390 (Bacchi et al. 2003),
and Perseus (Pedlar et al. 1990). All fluxes
are calculated in a consistent way from the fit procedure presented in
Paper II. 

The comparison between the radio power of mini halos and that of
the central cD galaxy, indicates that there is a weak tendency 
for more powerful mini-halos to host stronger central radio sources.
We recall that in a few clusters with cooling flows, the cD galaxy
can be a low power radio source or even radio quiet. 
This is indicative of a recurrent radio activity with a duty cycle of AGN
activity that is lower than the cooling time and lower than the radio 
mini-halo time life. 
Therefore, we do not expect a strong connection between cD and mini-halo
radio power, and the position  of Ophiuchus in the diagram
implies that this cluster is undergoing a
low radio activity phase of the central cD. We propose that this is 
important and that further studies and more robust statistical analyses 
are necessary to prove that
mini-halo emission is directly triggered by the central cD galaxy.

\begin{figure}
\begin{center}
\includegraphics[width=8cm]{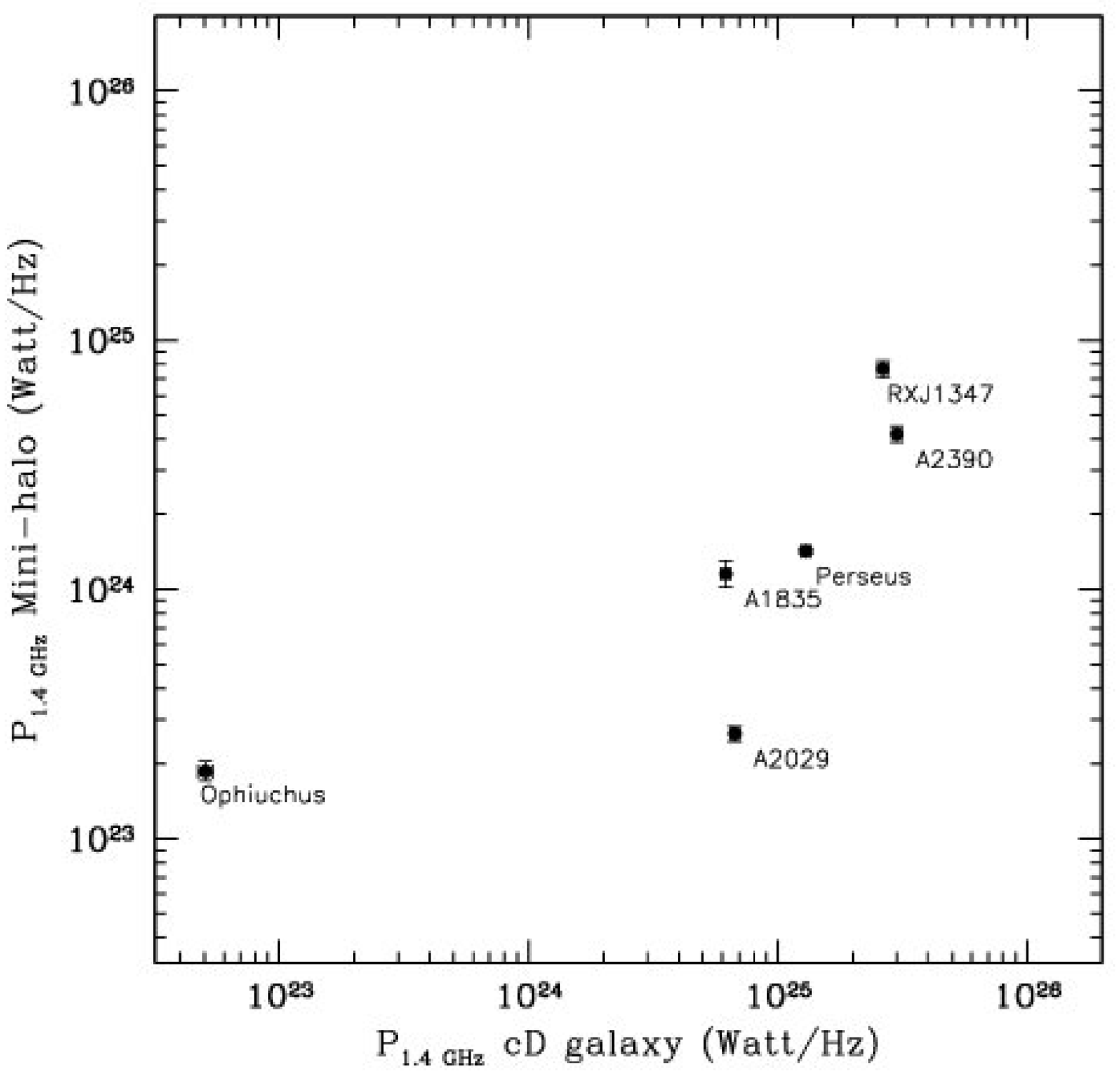}
\caption{Comparison between the radio power at 1.4 GHz of the 
mini-halos with that of the central cD galaxies, for the clusters analyzed
in this work. All the fluxes are derived from the fit procedure described in Paper II.}
\end{center}
\label{cD}
\end{figure}

\section{Conclusions}

Mini-halos in clusters are still poorly understood sources.  They are
a rare phenomenon, which have been found so far in only a few
clusters.  A larger number of mini-halo discoveries and more information about
their physical properties is necessary to discriminate between the
different mechanisms suggested for transferring energy to the
relativistic electrons that power the radio emission.

To search for new mini-halos, we have analyzed deep radio observations 
of A1068, A1413, A1650, A1835, A2029, and Ophiuchus, carried out 
with the Very Large Array at 1.4 GHz.

We have found that at the center of the clusters  A1835, A2029, and Ophiuchus, 
the dominant radio galaxy is surrounded by a diffuse 
low surface brightness mini-halo.
The relatively high resolution of our new images ensures 
that the detected diffuse emission is real and
not due to a blend of discrete sources.

We analyzed the interplay between the mini-halos and 
the cluster X-ray emission. We identified a similarity
between the shape of the radio mini-halo
emission and the cluster X-ray morphology of A2029, A1835, and Ophiuchus.
We note that, although all these clusters are considered to be 
relaxed systems, when analyzed in detail they are found to contain 
peculiar X-ray features at the cluster center,
which are indicative of a link
between the mini-halo emission and some minor merger activity.
Because of the large angular extension of Ophiuchus, it is 
possible to perform a 
point-to-point comparison of the radio and X-ray brightness distributions.
The close similarity between radio and X-ray structures in this
cluster is demonstrated by the correlation between these two parameters.
We fitted the data with a power law relation of the type
$I_{1.4 GHz}\propto I_X^{1.51\pm 0.04}$.

A hint of diffuse emission at the center of A1068 and A1413    
is present with low significance ($2-3 \sigma$), and
before classification as mini-halos, further investigation is needed.
In addition, in the field of view of one of
our observations, we report the serendipitous detection of 
a giant radio galaxy, located at 
RA=10$h$39$m$30$s$ DEC=39$^{\circ}$47$'$19$''$, 
of a total extension $\sim$1.6 Mpc in size.

Finally, the comparison between the radio power of mini halos and that of
the central cD galaxy, reveals that there is a weak tendency 
for the more powerful mini-halos to host stronger central radio sources.

Discriminating between the different scenarios proposed for  
mini-halo formation is difficult using present radio data.
We note that the radio-X ray connection in the Ophiuchus cluster may support 
secondary models. 
On the other hand, the analyses of the clusters studied here
appear to indicate that mini-halos are not a common 
phenomenon in relaxed system,
a result that is in closer agreement with primary models.
 
\begin{acknowledgements}
FG and MM thank the hospitality of the Harvard-Smithsonian Center 
for Astrophysics where most of this work was done.  
Support was provided by Chandra grants GO5-6123X and
GO6-7126X, NASA contract NAS8-39073, and the Smithsonian
Institution. This research was partially supported by ASI-INAF I/088/06/0 - 
High Energy Astrophysics and PRIN-INAF2005.
We are grateful to the referee Pasquale Mazzotta for very useful comments that 
improved this paper. We would like to thank Rossella Cassano and Chiara Ferrari 
for helpful discussions. The National Radio Astronomy Observatory (NRAO)
is a facility of the National Science Foundation, operated under
cooperative agreement by Associated Universities, Inc.
Based on photographic data obtained using The UK Schmidt Telescope. 
The UK Schmidt Telescope was operated by the Royal Observatory Edinburgh, 
with funding from the UK Science and Engineering Research Council, 
until 1988 June, and thereafter by the Anglo-Australian Observatory. 
Original plate material is copyright (c) of the Royal Observatory Edinburgh 
and the Anglo-Australian Observatory. 
The plates were processed into the present compressed digital form 
with their permission. The Digitized Sky Survey was produced at the 
Space Telescope Science Institute under US Government grant NAG W-2166.
This research has made use of the
NASA/IPAC Extragalactic Data Base (NED) which is operated by the JPL, 
California Institute of Technology, under contract with the National 
Aeronautics and Space Administration.

\end{acknowledgements}

\appendix
\section{Serendipitous detection of a giant radio galaxy south-west of A1068}

The left panel of Fig. A.1
shows a large field of view of the radio emission around A1068
overlaid on the archive XMM observation.
The XMM image is in the 0.2-5.0 keV band and has been convolved with 
a Gaussian of $\sigma =20''$.
The image shows a sky region south-west of A1068.
The cluster central emission is located in the top left corner 
of the figure.

At a projected distance of about 17$'$ from A1068, we detected 
an extraordinarily extended radio galaxy,
which does not belong to the cluster,
showing a wide-angle head-tail morphology. 
The core of the head-tail radio galaxy is located at
RA=10$h$39$m$30$s$ DEC=39$^{\circ}$47$'$19$''$
and is coincident with a bright galaxy at a 
redshift $z=0.0929$ (taken from the Sloan Survey).
At the distance of the head tail radio galaxy,
1$''$ corresponds to 1.71 kpc. Therefore, its total extension
reaches about 1.6 Mpc in size. 
The radio emission shows a clearly unresolved core, two jets
with some bright knots, and fading extended lobes.
The southern jet is brighter than the northern one. Moreover, 
it presents a well 
collimated shape that bends sharply to the east at about 
3$'$ ($\simeq$ 300 kpc) from the core. The morphology of the northern jet
is not as clear since it is only partially visible, although
it seems to bend more gradually than the southern jet. 
The difference in brightness and shape of the two jets
may indicate that the radio source appears shortened by 
projection effects.
On the north side, we see a diffuse low brightness lobe, while
the southern lobe is only partially detected. 
Because of its large extension, we may be missing flux 
from the entire source, and the flux density of the radio
galaxy calculated in our image must therefore be considered as a lower limit.
Its total flux density is $>$ 72 mJy, 
which corresponds to a power $L_{1.4GHz} > 1.5\times10^{24}$W/Hz.
From the NVSS, we estimated a flux density of $75\pm2$ mJy.

In the radio/X-ray overlay, we note an extended X-ray source, 
which may be a small cluster or a group hosting this giant radio galaxy.
The presence of a cluster is confirmed in the optical
band, since it belongs to the large new catalog of 
galaxy clusters (Koester et al. 2007) extracted from the Sloan Digital 
Sky Survey optical imaging data (York et al. 2000).
The interaction of the radio plasma with the thermal intracluster 
medium may explain its distorted morphology.

To identify the optical counterpart of the giant radio source,
the right panel of Fig. A.1 shows a zoomed image of the central
part of the radio galaxy overlaid on the optical image taken
from the POSS2 red plate.  The identification seems unambiguous.

\begin{figure*}
\begin{center}
\includegraphics[width=15cm]{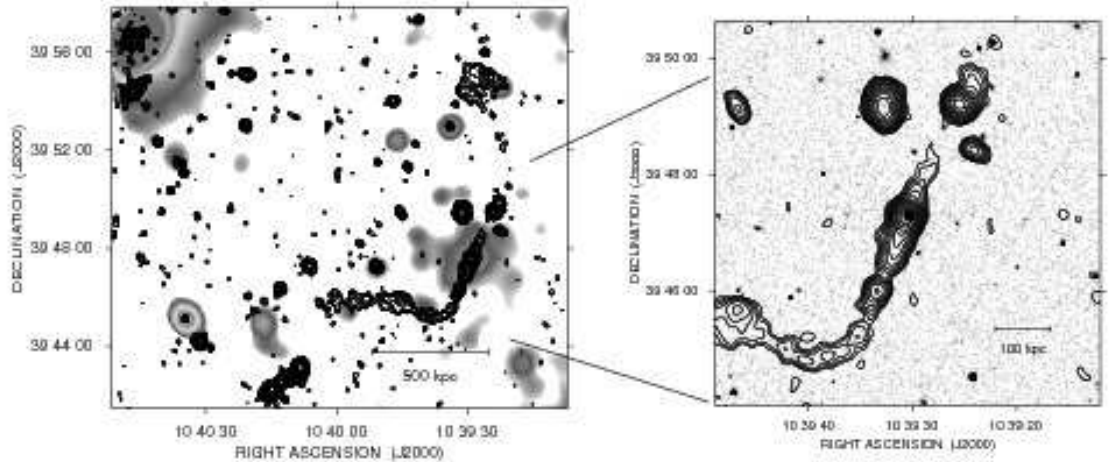}
\caption{
Total intensity radio contours on a large field of view
on south-west of the A1068 cluster, at 1.4 GHz with a
FWHM of 15$''\times 15''$ .
The first contour level is drawn at 0.065 mJy/beam
and the rest are spaced by a factor $\sqrt{2}$.
The sensitivity near the location of the giant radio galaxy (1$\sigma$) 
is 0.022 mJy/beam. 
On the left the radio contours are 
overlaid on the XMM image in the 0.2-5.0 keV, convolved with 
a Gaussian $\sigma = 20''$.
On the right a zoom of the radio contours are overlaid on the optical 
POSS2 image.
}
\label{giant}
\end{center}
\end{figure*}

\section{Serendipitous detection of an unusual radio galaxy in A1650}

Coincident with the galaxy
2MASX J12583829-0134290 (z=0.086217) that is at the position
RA=$12h58m38.2s$ DEC=$-01^{\circ}34'29''$, 
we detected an unusual radio galaxy. This radio galaxy
was also observed by Owen et al. (1993) in a 
1.4 GHz VLA survey of Abell clusters of galaxies.
The left panel of Fig. B.1 shows the radio emission 
with a resolution of 16$''$.
The morphology of the source appears to be a core with, 
to the north-west, a round structure with sharp edges.
The flux density of the overall structure is 135$\pm$4 mJy.
The right panel of Fig. B.1 shows the image taken from the
FIRST survey overlaid on the POSS2 image.
The higher resolution image shows that the round structure 
we detected at lower resolution is not due
to the radio emission from another galaxy.
It may be a NAT (narrow-angle tailed) source.
The faint extended emission visible at higher resolution
could be the ends of the two tails.
A similar source was detected in A3921 (Ferrari et al. 2006),
a NAT source associated with a galaxy, whose diffuse component is a partly
detached pair of tails from an earlier period of activity of the galaxy. 

The Chandra field of view does not extend to
the radio galaxy. We retrieved the larger field of view 
XMM observation from the archive but no diffuse X-ray emission
was detected at the location of the radio galaxy.

\begin{figure*}
\begin{center}
\includegraphics[width=15cm]{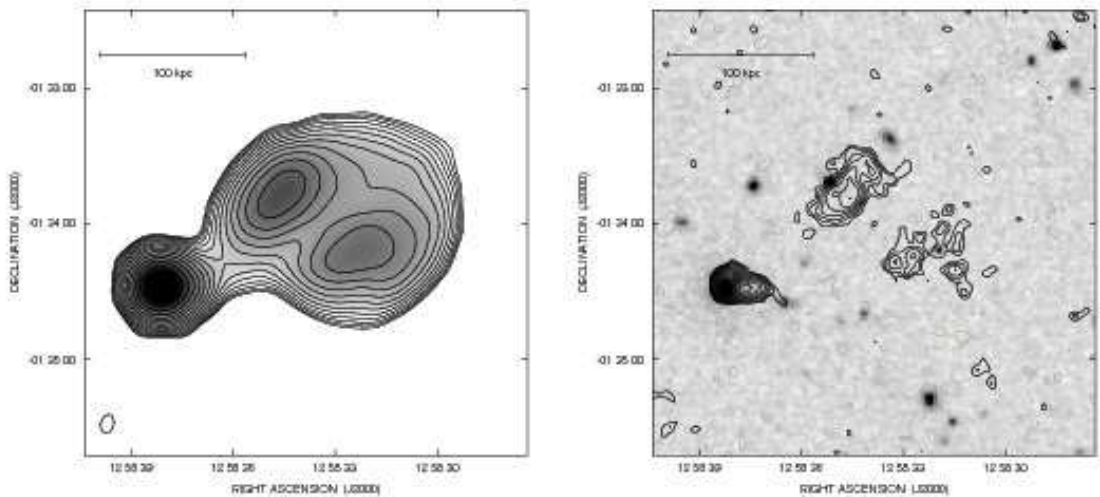}
\caption{
Left: total intensity radio contours to the north-west of the A1650 cluster, 
at 1.4 GHz with a FWHM of 16$''\times 16''$.
The first contour level is drawn at 0.13 mJy/beam
and the rest are spaced by a factor $\sqrt{2}$.
The sensitivity (1$\sigma$) is 0.045 mJy/beam.
Right: total intensity radio contours to the north-west of the A1650 cluster, 
at 1.4 GHz with a FWHM of 5.4$''\times 5.4''$ taken from the
FIRST survey.
The bottom two contour levels are drawn at $-$0.3 and 0.3 mJy/beam
and the rest are spaced by a factor $\sqrt{2}$.
The sensitivity of the FIRST survey (1$\sigma$) is 0.15 mJy/beam.
The contours of the radio intensity are overlaid on the optical 
POSS2 image.}
\end{center}
\label{A1650_ball}
\end{figure*}

\end{document}